\documentclass[11pt,reqno]{amsart}
\usepackage{amsfonts}
\usepackage{mathrsfs}
\usepackage{amsmath}
\usepackage{amssymb,amsfonts}

\newtheorem{thm}{Theorem}[section]
\newtheorem{lem}{Lemma}[section]
\newtheorem{defn}{Definition}[section]
\newtheorem{prop}{Proposition}[section]
\numberwithin{equation}{section}
\newtheorem{rmk}{Remark}[section]

\def\pf{{\textit {Proof:} }}

\newcommand{\mysection}[1]{\section{#1}\setcounter{equation}{0}}

\newfont{\bb}{msbm10 at 12pt}

\def\P{\mathcal P}

\newcommand{\bal}{\begin{aligned}}      \newcommand{\eal}{\end{aligned}}
\newcommand{\ba}{\begin{array}}      \newcommand{\ea}{\end{array}}
\newcommand{\bc}{\begin{center}}     \newcommand{\ec}{\end{center}}
\newcommand{\be}{\begin{enumerate}}  \newcommand{\ee}{\end{enumerate}}
\newcommand{\beq}{\begin{eqnarray}}  \newcommand{\eeq}{\end{eqnarray}}
\newcommand{\beQ}{\begin{eqnarray*}} \newcommand{\eeQ}{\end{eqnarray*}}
\newcommand{\bi}{\begin{itemize}}    \newcommand{\ei}{\end{itemize}}
\newcommand{\bt}{\begin{tabular}}    \newcommand{\et}{\end{tabular}}
\newcommand{\bdm}{\begin{displaymath}} \newcommand{\edm}{\end{displaymath}}

\def\qed{\hfill{Q.E.D.}\smallskip}

\newcommand{\ls}{\setlength{\baselineskip}{12pt}
                 \setlength{\parskip}{3mm}}

\begin{document}

\title[Positive energy theorem]{Positive energy theorem for asymptotically anti-de Sitter spacetimes with distributional curvature}

\author{Yaohua Wang$^{\dag}$}
\address[]{$^{\dag}$School of Mathematics and Information Science, Henan
University, Kaifeng, Henan 475004, PR China}
\email{wangyaohua@henu.edu.cn}
\author{Xiao Zhang$^{\flat}$}

\address[]{$^{\flat}$ Guangxi Center for Mathematical Research, Guangxi University, Nanning, Guanxi 530004, PR China}
\address[]{$^{\flat}$Institute of Mathematics, Academy of Mathematics and Systems Science, Chinese Academy of Sciences,
Beijing 100190, PR China and School of Mathematical Sciences, University of Chinese Academy of Sciences, Beijing 100049, P.R. China}

\email{xzhang@gxu.edu.cn, xzhang@amss.ac.cn}

\date{}

\begin{abstract}
We establish the positive energy theorem for weak asymptotically anti-de Sitter initial data
sets with distributional curvature under the weak dominant energy condition.


\end{abstract}

\maketitle \pagenumbering{arabic}

{\it {\small Dedicate to Professor Luen-Fai Tam on the occasion of his 70th birthday}}

\mysection{Introduction}\ls

The positive energy theorem plays a fundamental role in general relativity. When the cosmological constant is zero, the positive energy theorem was first proved by Schoen and Yau \cite{SY1,SY2}, then by Witten \cite{Wi,PT}.
When the cosmological constant is negative and spacetimes are asymptotically anti-de Sitter, the mathematical rigorous and complete proofs of the positive energy theorem were given by Wang, Chru\'sciel and Herzlich for asymptotically anti-de Sitter initial data sets with the trivial second fundamental form \cite{Wa, CH}, and by Chru\'{s}ciel, Maerten and Tod for them with the nontrivial second fundamental forms assuming the existence of center of AdS mass' coordinate systems \cite{M, CMT}. Finally, Wang, Xie and Zhang proved the positive energy theorem for general case \cite{WXZ}. Its extension involving electromagnetic fields was proved by Wang and Xu \cite{WX}.

It is an interesting question what is the weakest regularity of initial data sets for which the positive energy theorem holds. When the cosmological constant is zero and initial data sets have the trivial second fundamental forms, the positive energy theorem was proved by Miao for metrics which are Lipschitz along a hypersurface using the conformal deformation method \cite{Mi}. Similar results were also proved by Shi and Tam using Witten's spinor method together with an application to the proof of the positivity of the Brown-York quasi-local mass \cite{ST}, and by McFeron and Szkelyhidi using the Ricci flow method \cite{MS}.
Recently, Alaee and Yau provided the positive mass theorem with angular momentum and charges for axially symmetric initial
data sets with corners along a hypersurface \cite{AY}. When metrics have low-dimensional singular sets, Lee also used the conformal deformation method to proved the theorem \cite{L}. After that, Lee and LeFloch generalized all the above results to metrics with lower regularity $C^0\cap W^{1,n}_{-\tau}$ for $\tau\geq\frac{n-2}{2}$ \cite{LL}. Recently, Shibuya proved the following theorem which generalized Lee and LeFloch's result to the case of nontrivial second fundamental form \cite{S}.
\begin{thm}
Let $(M,g,h)$ be an n-dimensional $W^{1,n}_{-\tau}$ asymptotically flat data with $\tau > \frac{n-2}{2}$ and $P$ its generalized ADM $(n + 1)$-momentum . If this data has a spin structure and satisfies the dominant energy condition in distributional sense, then $P(U)$ is non-negative for any future-directed vector $U$ which is constant on the frame $\Phi$. In addition, if $P$ is zero, then the data has a globally parallel spinor frame with respect to the spacetime connection.
\end{thm}

The case of positive cosmological constant eventually can be reduced to the case of zero cosmological constant \cite{LXZ, LZ} and the relevant positive energy theorems hold automatically. But the case of negative cosmological constant is the most sophisticated case as it involves ten physical quantities. In this case Bonini and Qing proved the positive energy theorem when initial data sets have Lipschitz metrics along a hypersurface and the second fundamental form is trivial \cite{BQ}. In this paper, we proved it when the second fundamental form is nontrivial and generalized the main results of \cite{WXZ} to initial data sets with distributional curvatures.

\begin{thm}\label{pet}
Let $(M,g,h)$ be a 3-dimensional asymptotically anti-de Sitter initial data set of order $\tau >\frac{3}{2}$ with distributional curvature. Suppose that it satisfies the weak dominant energy condition (\ref{wdec}).
Let $E_0$ be the total energy, $c_{i}$, $c'_{i}$ and $J_{i}$ be the total momenta given in Definition \ref{def}. Then, for each end
\begin{equation}
\label{ineq}
E_0 \geq \sqrt{L^2 -2V^2 +2 \big(\max\{A^4 -L^2 V^2, 0\}\big)^\frac{1}{2}}.
\end{equation}
If $E_0 =0$ for some end, then $ {\widetilde R} _{ijij} ={\widetilde R} _{ijij} ^{AdS}, \,{\widetilde R} _{0jkl} = {\widetilde R} _{0jkl} ^{AdS}$ along $M$ in the sense of distribution.
\end{thm}

The paper is organized as follows:
In Section 2, we introduce weak asymptotically anti-de Sitter initial data sets with distributional curvature and weak dominant energy condition.
In Section 3, we study the existence and uniqueness of the Dirac equation for weak asymptotically anti-de Sitter initial data sets
satisfying weak dominant energy condition.
In Section 4, we show that the total energy-momentum can be defined for weak asymptotically anti-de Sitter initial data sets and prove the positive energy theorem under the weak dominant energy condition.

\mysection{Weak initial data sets}\ls

Let $(N, \widetilde{g})$ be a spacetime and $(M, g, h)$ be an initial data set where $M$ is a 3-dimensional spacelike hypersurface with the induced
Riemannian metric $g$ and the second fundamental form $h$. Let $\nabla$ and $\widetilde{\nabla}$
be the Levi-Civita connections of $g$ and $\widetilde{g}$ respectively. Let $\mathbb{S}$ be the locally
spinor bundle of $N$ and we still denote by $\mathbb{S}$ its restriction to $M$. Lift $\nabla$ and $\widetilde{\nabla}$
to $\mathbb{S}$ and denote the corresponding spin connections the same as $\nabla$ and $\widetilde{\nabla}$.
Fix a point $p\in M$ and an orthonormal basis $\{e_\alpha\}$ of $T_pN$ with
$e_0$ normal and $\{e_i\}$ tangent to $M$. Extend $\{e_\alpha\}$ to a local orthonormal frame in a neighborhood of
$p$ in $M$ such that $(\nabla^g_ie_j)_p=0$. Extend this to a local orthonormal frame $\{e_\alpha\}$ for $N$ with
$(\widetilde{\nabla}_0e_j)_p=0$. Then $(\widetilde{\nabla}_ie_j)_p=h_{ij}e_0$, $(\widetilde{\nabla}_ie_0)_p=h_{ij}e_j$.
Denote $H=\sum_{i}h_{ii}$, $|h|^2 =h_{ij} h^{ij}$.  There is a positive definite Hermitian metric $\langle\cdot,\cdot\rangle$ on $\mathbb{S}$,
with respect to which $e_i$ is skew-Hermitian and $e_0$ is Hermitian. Unless
Furthermore, $\nabla$ is compatible with $\langle\cdot,\cdot\rangle$, but $\widetilde{\nabla}$ is not.
The constraint equations give that
\begin{equation*}
\begin{aligned}
T_{00}=&\frac{1}{2}\big(R+H^2-|h| ^{2}\big)+3\kappa^2,\\
T_{0i}=&\nabla ^j \big(h_{ij}-g_{ij} H\big).
\end{aligned}
\end{equation*}
Define
\begin{equation*}
\widehat{\nabla}_i=\widetilde{\nabla}_i+\frac{\sqrt{-1}}{2}\kappa e_i, \quad \widehat D =\sum _{i=1} ^3 e_i \widehat{\nabla}_i.
\end{equation*}

Let $\breve{{\nabla}}$ and $\{\breve{e}_i\}$ be the Levi-Civita connection and frame of the hyperbolic metric
\beQ
\breve{g}=dr^2+\frac{\sinh^2(\kappa r)}{\kappa^2}\big(d\theta^2+\sin^2\theta d \psi^2\big)
\eeQ
respectively. Given 3-dimensional Riemannian manifold $M$, we assume that it equips with a smooth metric which is $\breve{g}$ on ends for simplicity. (Indeed, it is sufficient if there is an asymptotically hyperbolic complete metric of order $\tau>3$ in the sense of \cite{Z}). For abusing of notations, we denote this metric as $\breve{g}$ and $\breve{\rho}$ its distance function. Now we introduce the following function's weighed norm on $M$. For $p>0$,
\begin{equation*}
\| u \| _{L^{p}_{\alpha}}  =\int_{M}| u |^p e^{-\alpha p \kappa \breve{\rho} }d\mu_{\breve{g}}.
\end{equation*}
We define the weighted Sobolev spaces
\begin{equation*}
\aligned
L^{p}_{\alpha} &= \big\{u : \,\| u \| _{L^{p}_{\alpha}}<\infty   \big\},\\
W^{1,p} _{\alpha} & =\big\{u: \, \| u \| _{L^{p} _{\alpha}} +\| \nabla u \| _{L^{p}_{\alpha}} <\infty \big\}.
\endaligned
\end{equation*}

\begin{lem}\label{sobolev} The following propositions hold for weighted Sobolev spaces.
\bi
\item[(1)] $L^{p}_{\alpha}\subset L^{p}_{\beta}$ for $0<\alpha\leq\beta$. In particular, $L^{p}_{0}$ is the standard Sobolev space $L^{p}$ and $L^{p}_{\alpha}\subset L^{p}$ for $\alpha\leq 0$. \\
\item[(2)] If $f\in L^{p}_{\alpha}$, $g\in L^{q}_{\beta}$ for $p,q>0$ satisfying $\frac{1}{p}+\frac{1}{q}=1$, then $fg\in L^{1}_{\alpha+\beta}$.
 \item[(3)] If $f\in L^{p}_{\alpha}$ for $p>0$ and $\alpha<0$, then $f\in L^{p'}$ for $p'$ satisfying $\frac{\alpha p}{p-p'}<-2$.
\ei
\end{lem}
\pf
Since $e^{-\alpha p\kappa \breve{\rho}}\geq e^{-\beta p\kappa \breve{\rho}}$ for $0<\alpha\leq\beta$ and $e^{-\alpha p\kappa \breve{\rho}}=1$ for $\alpha=0$, then (1) holds. (2) can be proved by the H\"{o}lder inequality
\begin{eqnarray*}
\begin{aligned}
\int_{M} | f g | e^{-(\alpha+\beta) \kappa \breve{\rho}}d\mu_{\breve{g}}
\leq \left(\int_{M} | f | ^p e^{-\alpha p\kappa \breve{\rho}}d\mu_{\breve{g}}\right)^{\frac{1}{p}}\left(\int_{M} | g | ^q e^{-\beta q\kappa \breve{\rho}}d\mu_{\breve{g}}\right)^{\frac{1}{q}}.
\end{aligned}
\end{eqnarray*}
(3) can be proved by the H\"{o}lder inequality
\begin{eqnarray*}
\begin{aligned}
\int_{M} | f | ^{p'} d\mu_{\breve{g}}
\leq &\left(\int_{M} | f | ^p e^{-\alpha p\kappa \breve{\rho}}d\mu_{\breve{g}}\right)^{\frac{p'}{p}}\left(\int_{M}  e^{ \frac{\alpha p}{p-p'}\kappa \breve{\rho}}d\mu_{\breve{g}}\right)^{\frac{p-p'}{p}}\\
<&C\left(\int_{M} | f | ^p e^{-\alpha p\kappa \breve{\rho}}d\mu_{\breve{g}}\right)^{\frac{p'}{p}}.
\end{aligned}
\end{eqnarray*}
 \qed

In terms of the reference metric $\breve{g}$ together with its Levi-Civita connection $\breve{{\nabla}}$, we can write the scalar curvature of any metric $g$ as
\begin{eqnarray*}
R=\breve{\nabla}_k V^k+F,
\end{eqnarray*}
where
\begin{eqnarray*}
\begin{aligned}
V^k=&g^{ij}T^{k}_{ij}-g^{ik}T^{j}_{ji},\\
F=&\breve{R}-\breve{\nabla}_kg^{ij}T^{k}_{ij}+\breve{\nabla}_kg^{ik}T^{j}_{ji}+g^{ij}(T^{k}_{kl}T^{l}_{ij}-T^{k}_{jl}T^{l}_{ik}),\\
T^{k}_{ij}=&\frac{1}{2}g^{kl}(\breve{\nabla}_i g_{jl}+\breve{\nabla}_j g_{il}-\breve{\nabla}_l g_{ij}).
\end{aligned}
\end{eqnarray*}
Let set of vector $\vec{X}=(X^0, X^1,X^2,X^3)$
\beQ
\mathcal{U}=\Big\{ \vec{X}: X^\alpha \in W ^{1,2}\, \text{with compact support} \Big\}.
\eeQ
For any $\vec{X} \in \mathcal{U}$, we denote
\begin{equation*}
\begin{aligned}
{\mathcal{R}} (\vec{X})=&-\frac{1}{2}\int_M \breve{{\nabla}}_k\Big( X ^0\frac{d\mu_{g}}{d\mu_{\breve{g}}}\Big)V^k
d\mu_{\breve{g}}\\
&+\int_M\Big(\frac{1}{2}\big(F+H^2-|h| ^{2}\big)+3\kappa^2 \Big)X ^0 d\mu_{g}\\
&+\int_M \Big( h_{ij} -g_{ij} H \Big) \nabla ^j  X ^i d\mu_{g}.
\end{aligned}
\end{equation*}

\begin{defn}\label{def2}
Initial data set $(M,g,h)$ is weak asymptotically anti-de Sitter of order $\tau >\frac{3}{2}$ with distributional curvature if
\bi
\item[(1)] There is a compact set $K  \subset M$ such that $M
\setminus K$ is the disjoint union of a finite number of subsets (ends) $M_i$ and each $M_i$
is diffeomorphic to $\mathbb{R}^3 \setminus B_r$ with $B _r$ the closed ball of radius $r$;
\item[(2)] Under this diffeomorphism, $g-\breve{g}\in C^0\cap W^{1,2}_{-\tau+1}$, $h\in C^0\cap L^{2}_{-\tau+1}$;

\item[(3)] $\mathcal{R}(\vec{X}) < +\infty$ for any $\vec{X}\in W^{1,2}_{\frac{1}{2}}$.

\ei
\end{defn}

\begin{defn}\label{def1}
Let $(M,g,h)$ be a weak asymptotically anti-de Sitter initial data set of order $\tau >\frac{3}{2}$ with distributional curvature. It
satisfies the weak dominant energy condition if, for all future-directed, non-spacelike $\vec{X} \in \mathcal{U}$,
\beq
{\mathcal{R}} (\vec{X}) \geq 0. \label{wdec}
\eeq
\end{defn}

Throughout the paper we denote the cut-off function
\begin{equation*}
\chi_\rho =\left\{\aligned 0, &  \qquad r> \rho+\varepsilon, \\
1-\frac{r-\rho}{\varepsilon}, &  \qquad \rho < r \leq \rho +\varepsilon,\\
1, &  \qquad r \leq \rho.
\endaligned\right.
\end{equation*}

\mysection{Dirac equation}\ls

Recall (cf. \cite{WXZ}) that the anti-de Sitter spacetime equipped with the metric
\begin{equation*}
\widetilde{g}_{AdS}=-\cosh^2(\kappa r)dt^2+dr^2+\frac{\sinh^2(\kappa r)}{\kappa^2}\big(d\theta^2+\sin^2\theta d \psi^2\big)
\label{ads-metric}
\end{equation*}
is characterized by imaginary Killing spinors
\begin{eqnarray*}
\widetilde{\nabla}^{AdS}_X \Phi_0+\frac{\kappa \sqrt{-1}}{2}X\cdot\Phi_0=0.
\end{eqnarray*}
Fix the Clifford representation
\begin{eqnarray}
\begin{aligned}
\breve{e}_0 \mapsto \begin{pmatrix}\ &\ &1 &\ \\ \ &\ & \ &1\\1&\ &\ &\ \\ \ &1&\ &\ \end{pmatrix}, &\quad
\breve{e}_1 \mapsto \begin{pmatrix}\ &\ &-1 &\ \\ \ &\ & \ &1\\1&\ &\ &\ \\ \ &-1&\ &\ \end{pmatrix},\\
\breve{e}_2 \mapsto \begin{pmatrix}\ &\ &\ &1 \\ \ &\ & 1 &\ \\\ &-1 &\ &\ \\ -1 &\ &\ &\ \end{pmatrix}, &\quad
\breve{e}_3 \mapsto \sqrt{-1}\begin{pmatrix}\ &\ &\ &1 \\ \ &\ & -1 &\ \\\ &-1 &\ &\ \\ 1 &\ &\ &\ \end{pmatrix},
\label{repre}
\end{aligned}
\end{eqnarray}
then the imaginary Killing spinor $\Phi_0 ^\lambda$ takes the following form
\begin{equation}\label{ik}
\Phi_0 ^\lambda =\begin{pmatrix}u^+e^{\frac{\kappa r}{2}}+u^-e^{-\frac{\kappa r}{2}}\\v^+e^{\frac{\kappa
r}{2}}+v^-e^{-\frac{\kappa r}{2}}\\-\sqrt{-1}u^+e^{\frac{\kappa r}{2}}+\sqrt{-1}u^-e^{-\frac{\kappa r}{2}} \\
\sqrt{-1}v^+e^{\frac{\kappa r}{2}}-\sqrt{-1}v^-e^{-\frac{\kappa r}{2}}
\end{pmatrix},
\end{equation}
where
\begin{eqnarray*}\label{ik2}
\begin{aligned}
u^+ =&\Big(\lambda_1\cos\frac{\kappa t}{2}+\lambda_3\sin\frac{\kappa t}{2}\Big)
       e^{\frac{\sqrt{-1}}{2}\psi}\sin\frac{\theta}{2}\\
     &+\Big(\lambda_2\cos\frac{\kappa t}{2}+\lambda_4\sin\frac{\kappa t}{2}\Big)
       e^{\frac{-\sqrt{-1}}{2}\psi}\cos\frac{\theta}{2},\\
u^-=&\Big(-\lambda_1\sin\frac{\kappa t}{2}+\lambda_3\cos\frac{\kappa t}{2}\Big)
       e^{\frac{\sqrt{-1}}{2}\psi}\sin\frac{\theta}{2}\\
    &+\Big(-\lambda_2\sin\frac{\kappa t}{2}+\lambda_4\cos\frac{\kappa t}{2}\Big)
       e^{\frac{-\sqrt{-1}}{2}\psi}\cos\frac{\theta}{2},\\
v^+=&-\Big(-\lambda_1\sin\frac{\kappa t}{2}+\lambda_3\cos\frac{\kappa t}{2}\Big)
       e^{\frac{\sqrt{-1}}{2}\psi}\cos\frac{\theta}{2}\\
    &+\Big(-\lambda_2\sin\frac{\kappa t}{2}+\lambda_4\cos\frac{\kappa t}{2}\Big)
       e^{\frac{-\sqrt{-1}}{2}\psi}\sin\frac{\theta}{2}, \\
v^-=&-\Big(\lambda_1\cos\frac{\kappa t}{2}+\lambda_3\sin\frac{\kappa t}{2}\Big)
       e^{\frac{\sqrt{-1}}{2}\psi}\cos\frac{\theta}{2}\\
    &+\Big(\lambda_2\cos\frac{\kappa t}{2}+\lambda_4\sin\frac{\kappa t}{2}\Big)
       e^{\frac{-\sqrt{-1}}{2}\psi}\sin\frac{\theta}{2},
\end{aligned}
\end{eqnarray*}
and $\lambda_1$, $\lambda_2$, $\lambda_3$, $\lambda_4$ are four arbitrary complex numbers \cite{WXZ}.
The Killing spinor $\Phi_0 $ along $M$ is given by fixing $t$ in $\Phi_0 ^\lambda $.

Now we solve the Dirac equation for weak initial data sets with distributional curvature. Note that
$g=\breve g + a$ with $a\in W^{1,2}_{-\tau+1}$. Orthonormalizing $\breve e_i$ with respect to the metric
$g$ yields a gauge transformation
\beQ
\aligned
\mathcal{A}: SO(\breve g) &\rightarrow SO(g)\\
\breve e_i &\mapsto e_i
\endaligned
\eeQ
where $e_i=\big(\delta _{ik}-\frac{1}{2}a_{ik} +f\big) \breve e_k$, $f=O(a^2) \in W^{1,1}_{-2\tau+2}$.
It provides identifies between the two spin groups as well as the spinor bundles.

\begin{lem} Let $\ {\nabla}'=\mathcal{A}\circ\breve{{\nabla}}\circ\mathcal{A}^{-1}$. Then the following asymptotic formula holds
\begin{equation}\label{reminder1}
\sum_{j,\ j\neq i}Re\langle\phi,e_i\cdot
e_j\cdot({\nabla}_j-{\nabla}'_j)\phi\rangle=\frac{1}{4}\Big(\breve{\nabla^j}
g_{ij}-\breve{{\nabla}}_i
tr_{\breve g}(g)+b \Big)|\phi|^2.
\end{equation}
where $b\in L^{1}_{-2\tau+2}\cap L^{\frac{3}{2}}$.
\end{lem}
\pf The computation is standard (e.g. \cite{Z}). The term $b$ involve $(\breve{{\nabla}}a_{ik})a_{kj}\breve e_j$ and higher order terms.
Since $a\in W^{1,2}_{-\tau+1}$, we have $\breve{{\nabla}}a_{ik}\in L^2_{-\tau+1}$ and $a_{kj}\in L^2_{-\tau+1}$.
Lemma \ref{sobolev} (2) gives $(\breve{{\nabla}}a_{ik})a_{kj} \in L^{1}_{-2\tau+2}$, and
Lemma \ref{sobolev} (3) gives $\breve{{\nabla}}a_{ik}\in L^{\frac{3}{2}}$.
Hence $(\breve{{\nabla}}a_{ik})a_{kj} \in L^{\frac{3}{2}}$ as $a_{ik}\in C^0$.
The lemma follows from these estimates. \qed

We extend the imaginary Killing spinors $\Phi_0$ smoothly to $M$. Then $\overline{\Phi}_0=\mathcal{A}\Phi_0$ are their pullback to the metric $g$.  (For sake of simplicity, we omit $\lambda$ here.)
Let $\widehat{\nabla} '_X={\nabla}'_X+\frac{\sqrt{-1}}{2}\kappa X $. It is straightforward that
\begin{eqnarray}\label{reminder2}
\begin{aligned}
\widehat{\nabla} '_j\overline{\Phi}_0=&\mathcal{A}\Big(-\frac{\sqrt{-1}}{2}\kappa e_j\cdot\Phi_0\Big)+\frac{\sqrt{-1}}{2}\kappa e_j\cdot\overline{\Phi}_0\\
=&\mathcal{A}\Big(-\frac{\sqrt{-1}}{2}\kappa \big(\delta _{ij}-\frac{1}{2}a_{ij} +f\big) \breve e_i \cdot\Phi_0\Big)+\frac{\sqrt{-1}}{2}\kappa e_j\cdot\overline{\Phi}_0\\
=&\frac{\sqrt{-1}}{4}\kappa a_{jk}e_k\cdot\overline{\Phi}_0+c\overline{\Phi}_0
\end{aligned}
\end{eqnarray}
for some $c\in L^{1}_{-2\tau+2}$.

Let $H^1(\mathbb{S})$ be the set of spinors with finite $W^{1,2}$ norm. For any $\phi \in H^1(\mathbb{S})$, we define the associated vector $\vec{X} _\phi$ where
\beQ
X _\phi ^0 = \chi_\rho\langle \phi, \phi \rangle, \quad X _\phi ^i = \chi_\rho\langle \phi, e_0 e_i \phi \rangle.
\eeQ
Note that $\vec{X} _\phi $ is always future-directed, non-spacelike (eg. see the proof of Lemma 1 \cite{HZ}).

Recall that for smooth $g$, $h$ and compactly supported smooth spinor $\phi$, the Weitzenb\"{o}ck formula gives
\begin{eqnarray*}
\begin{aligned}
\int_M\langle\widehat\nabla\psi,\widehat\nabla\phi\rangle -\langle\widehat D\psi,\widehat D\phi\rangle+ \langle\psi,\widehat{\mathcal{R}}\phi\rangle
=0
\end{aligned}
\end{eqnarray*}
where $\widehat{\mathcal{R}}=\frac{1}{2}(T_{00}-T_{0i}e_0 e_i)$.

By applying the density argument, similar to the discussions of Proposition 3.2 in \cite{LL} and Lemma 12 in \cite{S}, we can derive the following Weitzenb\"{o}ck formula in the distributional case.
\begin{lem}\label{WI}
Let $(M,g,h)$ be a weak asymptotically anti-de Sitter initial data set of order $\tau >\frac{3}{2}$ with distributional curvature. Then we have for any $\phi \in H^1(\mathbb{S})$,
\begin{eqnarray*}
\int_M \left(\langle -\widehat D\left(\chi_\rho \phi\right), \widehat D\phi \rangle+\langle\widehat\nabla\left(\chi_\rho \phi\right),\widehat\nabla\phi\rangle\right)+  \mathcal{R} (\vec{X} _\phi)=0.
\end{eqnarray*}
\end{lem}

The following lemma can be proved using Lemma \ref{WI}
\begin{lem}\label{convex}
Let $\widehat{\nabla}^\ast$, $\widehat{D}^\ast$ be the adjoint operators of $\widehat{\nabla}$, $\widehat{D}$ respectively. If the weak dominant energy condition holds, then, for any $\phi \in H^1(\mathbb{S})$,
\beQ
\int_M |\widehat D\phi|^2 \geq \int_M |\widehat\nabla\phi|^2, \qquad \int_M |\widehat D^\ast\phi|^2 \geq \int_M |\widehat\nabla^\ast\phi|^2.
\eeQ
\end{lem}
\pf
By Lemma \ref{WI},
\begin{eqnarray*}
\begin{aligned}
-\int_M e_i(\chi_\rho)\langle\phi,(\widehat\nabla_i+e_i\widehat D)\phi \rangle
=&\int_M  \left(-\chi_\rho|\widehat D\phi|^2+\chi_\rho|\widehat\nabla\phi|^2\right) + \mathcal{R} (\vec{X} _\phi)\\
\geq &-\int_M\chi_\rho|\widehat D\phi|^2+ \int_M \chi_\rho|\widehat\nabla\phi|^2.
\end{aligned}
\end{eqnarray*}
Since the left hand side goes to zero as $\rho\rightarrow\infty$, we obtain
\begin{eqnarray}\label{A}
\int_M |\widehat D\phi|^2 \geq   \int_M |\widehat\nabla\phi|^2.
\end{eqnarray}
Similarly,
\begin{eqnarray}\label{B}
\int_M |\widehat D^\ast\phi|^2 \geq \int_M |\widehat\nabla^\ast\phi|^2.
\end{eqnarray}
\qed

\begin{prop}\label{Dirac}
Let $(M,g,h)$ be a weak asymptotically anti-de Sitter initial data set of order $\tau >\frac{3}{2}$ with distributional curvature. Suppose that the weak dominant energy condition holds. Then there exists a unique spinor $\Phi_1$ in $H^1(\mathbb{S})$  such that
\begin{eqnarray*}
\widehat D\big(\Phi_1+\overline{\Phi}_0 \big)=0.
\end{eqnarray*}
\end{prop}
\pf The proof is similar to that of Lemma 4.2 in \cite{Z}. Define the bilinear form
\beQ
\mathcal{B}(\phi,\psi)=\int_M\langle\widehat{D}^\ast\phi,\widehat{D}^\ast\psi\rangle
\eeQ
on $H^1(\mathbb{S})$. If the weak dominant energy condition holds, then (\ref{A}) implies that $\mathcal{B}(\cdot,\cdot)$ is coercive. Since $\widehat D \overline{\Phi}_0 \in L^2(\mathbb{S})$, $\widehat\nabla \overline{\Phi}_0 \in L^2(\mathbb{S})$, there exists a spinor $\phi_1 \in H^1(\mathbb{S})$ such that, for any $\psi\in H^1(\mathbb{S})$,
\beQ
\int_M\langle\widehat D^\ast\phi_1,\widehat D^\ast\psi\rangle=-\int_M\langle\widehat D\overline{\Phi}_0,\psi\rangle.
\eeQ
Let $\Phi_1=\widehat D^\ast\phi_1$, and $\Phi=\Phi_1+\overline{\Phi}_0$, then $\Phi_1 \in L^2(\mathbb{S})$ and $$\int_M\langle\Phi,\widehat D^\ast\psi\rangle=0$$
for any $\psi\in H^1(\mathbb{S})$.
Let $\phi_j$ be a sequence of $H^1(\mathbb{S})$ spinors
converging to $\Phi_1$ in $L^2(\mathbb{S})$. For any $\psi\in H^1(\mathbb{S})$, we get
\begin{eqnarray*}
\lim_{j\rightarrow\infty}\langle\widehat D(\phi_j+\overline{\Phi}_0),\psi \rangle_{L^2}=\lim_{j\rightarrow\infty}\langle(\phi_j+\overline{\Phi}_0),\widehat D^\ast\psi \rangle_{L^2}=\langle\Phi,\widehat D^\ast\psi \rangle_{L^2}
=0.
\end{eqnarray*}
Thus $\widehat D(\phi_j+\overline{\Phi}_0)$ converges to $0$ in the weak $L^2(\mathbb{S})$ topology and $|\widehat D\phi_j|_{L^2}$ is bounded independently of $j$. This implies that $|\phi_j|_{H^1}$ is bounded independently of $j$ and $\phi_j$
converging to $\Phi_1$ weakly in $H^1(\mathbb{S})$. Therefore,  $\widehat D\Phi \in L^2(\mathbb{S})$ and $\widehat D\Phi=0$.

Now we prove the uniqueness. If there exist $\Phi_i\in H^1(\mathbb{S})$ such that $
\widehat D(\Phi_i+\overline{\Phi}_0)=0$ for $i=1,2$.
Let $\Psi=\Phi_1-\Phi_2$, then
$\widehat D\Psi=0$
and $\Psi\in H^1(\mathbb{S})$. The inequality (\ref{A}) implies that
$\widehat \nabla \Psi=0$.
Denote $M_R$ a domain of one end satisfying $\breve{\rho}>R$ for some large $R$. Denote $d\breve{\omega}$ the volume form of the standard round sphere $S^2$. Let $A$ be a positive constant which is determined later. Since $\Psi\in L^2(\mathbb{S})$, we have
\beQ
\int_{S^2} \left(\int_r^{+\infty} \chi _{\breve{\rho}} |\Psi|^2 e^{-(A+2\kappa)}\sqrt{g}d\breve{\rho}\right)d\breve{\omega} <+\infty
\eeQ
for any positive constant $A$. Therefore we obtain
\begin{eqnarray*}
\begin{aligned}
&\int_{S^2} \left(\int_r^{+\infty} \chi _{\breve{\rho}} |\Psi|^2 e^{-(A+2\kappa)\breve{\rho}}\sqrt{g}d\breve{\rho}\right)d\breve{\omega} \\
\leq&-\frac{C}{A}\int_{S^2} \left(\int_r^{+\infty}  \chi _{\breve{\rho}} |\Psi|^2 d(e^{-A\breve{\rho}})\right)d\breve{\omega} \\
\leq&\frac{C}{A}\int_{S^2} \left(\int_r^{+\infty} ( 2\chi _{\breve{\rho}} |\langle\nabla_{\breve{e}_1}\Psi,\Psi \rangle| e^{-A\breve{\rho}}+ \chi_{\breve{\rho}} '|\Psi|^2 e^{-A\breve{\rho}})d\breve{\rho}\right)d\breve{\omega} \\
\leq&\frac{2C}{A}\int_{S^2} \left(\int_r^{+\infty} \chi _{\breve{\rho}} |\langle \frac{1}{2}h(\breve{e}_1, e_j)e_0\cdot e_j\cdot\Psi  -\frac{\kappa\sqrt{-1}}{2}\breve{e}_1\cdot\Psi ,\Psi \rangle| e^{-A\breve{\rho}}d\breve{\rho}\right)d\breve{\omega} \\
&+\frac{C}{A}\int_{S^2} \left(\int_r^{+\infty}  C_1\chi _{\breve{\rho}} |\Psi|^2 e^{-A\breve{\rho}}d\breve{\rho}\right)d \breve{\omega} \\
\leq &\frac{C_2|h |_{C^0}+C_3\kappa+C_1}{A}\int_{S^2} \left(\int_r^{+\infty} \chi _{\breve{\rho}}  |\Psi|^2 e^{-A\breve{\rho}}d\breve{\rho}\right)
d\breve{\omega}\\
\leq &\frac{\bar{C}}{A}\int_{S^2} \left(\int_r^{+\infty} \chi _{\breve{\rho}} |\Psi|^2 e^{-(A+2\kappa)\breve{\rho}}\sqrt{g}d\breve{\rho}\right)d\breve{\omega}.
\end{aligned}
\end{eqnarray*}
The inequality gives, by choosing $A>\bar{C}$, that
\beQ
\int_{M}  \chi _{\breve{\rho}} |\Psi|^2e^{-A\breve{\rho}}\sqrt{g}d\breve{\rho}d\breve{\omega}=0.
\eeQ
Hence $\Psi=0$ on $M_R$. By Lemma 9.1 of \cite{BC}, we obtain $\Psi=0$ on $M-M_R$ and the proof of this proposition is complete.
\qed

The proof of the uniqueness in the Proposition \ref{Dirac} also implies the following proposition.

\begin{prop}\label{Dirac2}
Let $(M,g,h)$ be a weak asymptotically anti-de Sitter initial data set of order $\tau >\frac{3}{2}$ with distributional curvature. Suppose that $\Phi _{0\alpha}$ are linearly independent Killing spinors and $\Phi_{1\alpha} \in H^1(\mathbb{S})$. If $$\widehat \nabla \Phi _\alpha =0,\qquad
\Phi _\alpha = \Phi _{1\alpha} + \overline{\Phi}_{0\alpha},$$ then $\Phi_\alpha$ are linearly independent everywhere in $M$.
\end{prop}

\mysection{Total energy-momentum and proof of Theorem \ref{pet}}\ls

In this section we first define the total energy-momentum for weak initial data sets. Denote
\begin{eqnarray*}
\mathcal{E}_i=\breve{\nabla}^j g_{ij}-\breve
{{\nabla}}_itr_{\breve{g}}(g)-\kappa(a_{1i}-g_{1i}tr_{\breve{g}}(a)),\quad
\mathcal{P}_{ki}=h_{ki}-g_{ki}tr_{\breve{g}}(h).
\end{eqnarray*}
Let $U_{\alpha \beta}$ be the restrictions of the Killing vectors on the $t$-slice (cf. Appendix A \cite{WXZ}).
Given $\varepsilon>0, \rho>0$, we introduce the following ten pre total energy-momentum
\begin{eqnarray*}\label{e-m}
\begin{aligned}
E _0(\varepsilon, \rho)=&\frac{\kappa}{16\pi}\frac{1}{\varepsilon}\int_{\rho<r<\rho+\varepsilon}\mathcal{E}_1 U_{40}^{(0)}d\mu_{\breve{g}},\\
{c}_{i}(\varepsilon, \rho)=&\frac{\kappa}{16\pi}\frac{1}{\varepsilon}\int_{\rho<r<\rho+\varepsilon}\mathcal{E}_1 U_{i4}^{(0)}d\mu_{\breve{g}}+\frac{\kappa}{8\pi}\sum_{j=2}^{3}\frac{1}{\varepsilon}\int_{\rho<r<\rho+\varepsilon}\mathcal{P}_{j1}U_{i4}^{(j)} d\mu_{\breve{g}},\\
{c}'_{i}(\varepsilon, \rho)=&\frac{\kappa}{16\pi}\frac{1}{\varepsilon}\int_{\rho<r<\rho+\varepsilon}\mathcal{E}_1 U_{i0}^{(0)}d\mu_{\breve{g}}+\frac{\kappa}{8\pi}\sum_{j=2}^{3}\frac{1}{\varepsilon}\int_{\rho<r<\rho+\varepsilon}\mathcal{P}_{j1}U_{i0}^{(j)}
      d\mu_{\breve{g}} ,\\
{J}_{i}(\varepsilon, \rho)=&\frac{\kappa}{8\pi}\sum_{j=2}^{3}\frac{1}{\varepsilon}\int_{\rho<r<\rho+\varepsilon}\mathcal{P}_{j1}V_{i}^{(j)}
d\mu_{\breve{g}}.
\end{aligned}
\end{eqnarray*}
With respect to Clifford multiplication (\ref{repre}), the pre energy-momentum matrix is
\beq \label{Q}
\begin{aligned}
 {\bf Q}(\varepsilon, \rho) =\begin{pmatrix}
 P  (\varepsilon, \rho)        &     W (\varepsilon, \rho)\\
 \overline{W}^t (\varepsilon, \rho) &    \hat{P} (\varepsilon, \rho)
 \end{pmatrix},
\end{aligned}
\eeq
where
\beQ
\begin{aligned}
P(\varepsilon, \rho)&=\begin{pmatrix}
 E_0(\varepsilon, \rho)-c_3(\varepsilon, \rho)             &      c_1(\varepsilon, \rho)-\sqrt{-1}c_2(\varepsilon, \rho)\\
 c_1(\varepsilon, \rho)+\sqrt{-1}c_2(\varepsilon, \rho)    &      E_0(\varepsilon, \rho)+c_3(\varepsilon, \rho)
 \end{pmatrix},\\
\hat{P}(\varepsilon, \rho)&=\begin{pmatrix}
 E_0(\varepsilon, \rho)+c_3 (\varepsilon, \rho)            &      -c_1(\varepsilon, \rho)+\sqrt{-1}c_2(\varepsilon, \rho)\\
 -c_1(\varepsilon, \rho)-\sqrt{-1}c_2 (\varepsilon, \rho)  &      E_0(\varepsilon, \rho)-c_3(\varepsilon, \rho)
 \end{pmatrix},\\
W(\varepsilon, \rho)&=\begin{pmatrix}
 w_1 (\varepsilon, \rho) &     w_2 ^+ (\varepsilon, \rho)\\
 w_2 ^- (\varepsilon, \rho) &    -w_1 (\varepsilon, \rho)
 \end{pmatrix},\\
w_1(\varepsilon, \rho) &= c'_{3}(\varepsilon, \rho)-\sqrt{-1}J_{3}(\varepsilon, \rho),\\
w_2 ^\pm (\varepsilon, \rho)& = -c'_{1}(\varepsilon, \rho) \pm J_{2}(\varepsilon, \rho)\pm \sqrt{-1}(c_{2}'(\varepsilon, \rho)\pm J_{1}(\varepsilon, \rho)).
\end{aligned}
\eeQ

\begin{lem}\label{IWI1}
Let $(M,g,h)$ be an asymptotically anti-de Sitter initial data set of order $\tau >\frac{3}{2}$ with distributional curvature. Suppose it satisfies the weak dominant energy condition. Let $\phi$ be the solution of the Dirac-type equation
$\widehat D\phi=0$ obtained in Proposition \ref{Dirac}. Then
\begin{eqnarray}
\begin{aligned}\label{IWI}
8 \pi\vec{\lambda}\, {\bf Q}(\varepsilon, \rho)\,\vec{\lambda} ^t+\frac{1}{2\varepsilon}\int_{\rho<r<\rho+\varepsilon}
          O(L^1) d\mu_{\breve{g}} =\int_M \chi_\rho|\widehat\nabla\phi|^2 + \mathcal{R} (\vec{X} _\phi).
          \end{aligned}
\end{eqnarray}

\end{lem}
\pf Since $\widehat D\phi=0$ and
$$\int_M\langle\widehat\nabla\left(\chi_\rho \phi\right),\widehat\nabla\phi\rangle+  \mathcal{R} (\vec{X} _\phi) =0,$$
we obtain,
\begin{eqnarray*}
\begin{aligned}
& \int_M \chi_\rho|\widehat\nabla\phi|^2 + \mathcal{R} (\vec{X} _\phi)\\
=&-\int_M\langle (\widehat\nabla \chi_\rho) \phi,\widehat\nabla\phi\rangle\\
=&\frac{1}{4\varepsilon}\int_{\rho<r<\rho+\varepsilon}
         (\breve{\nabla}^j g_{1j}-\breve{{\nabla}}_1tr_{\breve{g}}(g))|\Phi_0|^2d\mu_{\breve{g}}\\
&+\frac{1}{4\varepsilon}\int_{\rho<r<\rho+\varepsilon}
         \kappa(a_{k1}-g_{k1}tr_{\breve{g}}(a))\langle\Phi_0,\sqrt{-1}\breve{e}_k\cdot\Phi_0 \rangle d\mu_{\breve{g}}\\
&-\frac{1}{2\varepsilon}\int_{\rho<r<\rho+\varepsilon}
        (h_{k1}-g_{k1}tr_{\breve{g}}(h))\langle\Phi_0,\breve{e}_0\cdot\breve{e}_k\cdot\Phi_0\rangle d\mu_{\breve{g}}\\
&+\frac{1}{2\varepsilon}\int_{\rho<r<\rho+\varepsilon} Err\,\, d\mu_{\breve{g}}\\
=&8 \pi\vec{\lambda}\, {\bf Q}(\varepsilon, \rho)\,\vec{\lambda} ^t+\frac{1}{2\varepsilon}\int_{\rho<r<\rho+\varepsilon}
          Err\,\, d\mu_{\breve{g}}.
\end{aligned}
\end{eqnarray*}

Now we show that the error term $Err$ belongs to $L^1$. The error term contains the following three types of spinors:

(1) $b|\overline{\Phi}_0|^2$ where $b\in L^{1}_{-2\tau+2}$. In this case
\begin{eqnarray*}
\begin{aligned}
         \int_M b|\overline{\Phi}_0|^2d\mu_{\breve{g}}
         \leq C_1\int_Mbe^{\kappa \breve{\rho}}d\mu_{\breve{g}} \leq C\int_Mbe^{(2\tau-2)\kappa \breve{\rho}}d\mu_{\breve{g}}<\infty.
          \end{aligned}
\end{eqnarray*}

(2) $h_{k1}\langle\Phi_0,\breve{e}_0\cdot\breve{e}_k\cdot\Phi_1\rangle$. Since $\Phi_1\in L^2$, $h_{k1}\in L^{2}_{-\tau+1}$,
\begin{eqnarray*}
\begin{aligned}
         \left(\int_M |h_{k1}\langle\Phi_0,\breve{e}_0\cdot\breve{e}_k\cdot\Phi_1\rangle|d\mu_{\breve{g}}\right)^2
         &\leq C_1\int_M|h_{k1}|^2|\Phi_0|^2d\mu_{\breve{g}}\int_M|\Phi_1|^2d\mu_{\breve{g}}\\
         &\leq C_2\int_M|h_{k1}|^2|\Phi_0|^2d\mu_{\breve{g}}\\
         &\leq C\int_M|h_{k1}|^2e^{(2\tau-2)\kappa \breve{\rho}}d\mu_{\breve{g}}<\infty.
\end{aligned}
\end{eqnarray*}

(3) $c|\Phi_1|^2$ where $c\in C^0$ or $c\in L^{\frac{3}{2}}$. If $c\in C^0$, as $\Phi_1\in L^2$,
\begin{eqnarray*}
\begin{aligned}
         \int_M c|\Phi_1|^2d\mu_{\breve{g}}
         \leq C_1\int_M|\Phi_1|^2d\mu_{\breve{g}}<\infty.
          \end{aligned}
\end{eqnarray*}
If $c\in L^{\frac{3}{2}}$, as $\Phi_1\in W^{1,2}$, the Sobolev embedding theorem implies $\Phi_1\in L^{6}$. Using H$\ddot{o}$lder inequality, we obtain
\begin{eqnarray*}
\begin{aligned}
         \int_M c|\Phi_1|^2d\mu_{\breve{g}}
         \leq \left(\int_M |c|^{\frac{3}{2}}d\mu_{\breve{g}}\right)^{\frac{2}{3}}\left(\int_M|\Phi_1|^6d\mu_{\breve{g}}\right)^{\frac{1}{3}}<\infty.
          \end{aligned}
\end{eqnarray*}
Therefore, (\ref{IWI}) follows. \qed

\begin{prop}\label{lim}
Let $(M,g,h)$ be a weakly asymptotically anti-de Sitter initial data set of order $\tau >\frac{3}{2}$ with distributional curvature. Then the following limits exist and they do not depend on $\varepsilon$
\beQ
\begin{aligned}
E_0=\lim _{\rho \rightarrow \infty} E _0(\varepsilon, \rho),\,\,c_{i}=\lim _{\rho \rightarrow \infty} c_{i}(\varepsilon, \rho),\,\,
c'_{i}=\lim _{\rho \rightarrow \infty} c'_{i}(\varepsilon, \rho), \,\,J_{i}= \lim _{\rho \rightarrow \infty} J _{i}(\varepsilon, \rho).
\end{aligned}
\eeQ
\end{prop}
\pf Recall that the following Weitzenb\"{o}ck formula holds
\begin{eqnarray*}
\int_M\left( \langle- \widehat D\left(\chi_\rho \phi\right), \widehat D\phi \rangle+\langle\widehat\nabla\left(\chi_\rho \phi\right),\widehat\nabla\phi\rangle\right)+  \mathcal{R} (\vec{X} _\phi)
=0.
\end{eqnarray*}
Inserting $\overline{\Phi}_0 $ (See Section 3) into the equation above, we obtain
\begin{eqnarray*}
\int_M \left(\langle -\widehat D\left(\chi_\rho \overline{\Phi}_0\right), \widehat D\overline{\Phi}_0 \rangle+\langle\widehat\nabla\left(\chi_\rho \overline{\Phi}_0\right),\widehat\nabla\overline{\Phi}_0\rangle\right)+  \mathcal{R} (\vec{X} _{\overline{\Phi}_0})
=0.
\end{eqnarray*}
Argument similar to Lemma \ref{IWI1} shows that
\begin{eqnarray*}
\begin{aligned}
&\int_M \left(-\chi_\rho |\widehat D\overline{\Phi}_0|^2+\chi_\rho|\widehat\nabla\overline{\Phi}_0|^2\right)+ \mathcal{R} (\vec{X} _{\overline{\Phi}_0})\\
=&\int_M-\langle (\widehat\nabla \chi_\rho) \overline{\Phi}_0,\widehat\nabla\overline{\Phi}_0\rangle+\langle\widehat
D(\chi_\rho) \overline{\Phi}_0, \widehat D\overline{\Phi}_0 \rangle\\
=& 8 \pi\vec{\lambda}\, {\bf Q}(\varepsilon, \rho)\,\vec{\lambda} ^t+\frac{1}{2\varepsilon}\int_{\rho<r<\rho+\varepsilon}
          O(L^1) d\mu_{\breve{g}}.
          \end{aligned}
\end{eqnarray*}
We will prove the result for $E_0$, other qualities may be discussed similarly.
Taking $\vec{\lambda}_1=(1,0,0,0)$ and $\vec{\lambda}_2=(0,1,0,0)$ and let $\phi_i$  be $\overline{\Phi}_0$ corresponding for $\vec{\lambda}_i$, then taking summation, one derives that
\begin{eqnarray*}
\begin{aligned}
         &\int_M \chi_\rho(|\widehat\nabla\phi_1|^2 +|\widehat\nabla\phi_2|^2 -\chi_\rho |\widehat D\phi_1|^2 -\chi_\rho |\widehat D\phi_2|^2) + \mathcal{R} (\vec{X} _{\phi_1})+ \mathcal{R} (\vec{X} _{\phi_2}) \\
          =&16 \pi E _0(\varepsilon, \rho)+\frac{1}{2\varepsilon}\int_{\rho<r<\rho+\varepsilon}
          O(L^1) d\mu_{\breve{g}}.
          \end{aligned}
\end{eqnarray*}
Since $\widehat\nabla\phi_i,\widehat D\phi_i\in L^2$,  and the limit of $\mathcal{R} (\vec{X} _{\phi_i})$ is finite as $\rho\rightarrow \infty$, Lebesgue¡¯s
dominated convergence theorem implies $E _0(\varepsilon, \rho)$ convergence to a finite limit as $\rho\rightarrow \infty$ which is independent of $\varepsilon$. Note that the Killing spinor $\Phi _0$ is pulled back on the end and extended to the inside of $M$, and
$E _0$ is obtained by taking $ \rho\rightarrow \infty$ lying on the end, thus it does not
depend on how one extends the Killing spinor to the inside of $M$ to get $\overline{\Phi}_0$. \qed

\begin{defn}\label{def}
Let $(M,g,h)$ be a weak asymptotically anti-de Sitter initial data set of order $\tau >\frac{3}{2}$ with distributional curvature. $E_0$ is defined as the total energy, $c_{i}$, $c'_{i}$ and $J_{i}$ are defined as the total momenta.
\end{defn}

As in \cite{WXZ}, we denote ${\bf c} =(c_1, c_2, c_3)$, ${\bf c}' =(c' _1, c' _2, c' _3)$, ${\bf J} =(J_1, J_2, J_3)$ and
\beq\label{3constants}
\begin{aligned}
L=&\big(|{\bf c}|^2 +|{\bf c}'|^2 +|{\bf J}|^2 \big)^{\frac{1}{2}},\\
A=&\big(|{\bf c} \times {\bf c}' |^2+|{\bf c} \times {\bf J}|^2 +| {\bf c}'\times {\bf J} |^2\big)^{\frac{1}{4}},\\
V=&\big(\varepsilon_{ijl}c_i c_{j}' J_l \big) ^{\frac{1}{3}},
\end{aligned}
\eeq
where $2L$, $2A^2$ and $V^3$ are the (normalized) length, surface area and volume of the parallelepiped spanned by
${\bf c}$, ${\bf c}'$ and ${\bf J}$.

{\em Proof of Theorem \ref{pet}}. From Proposition \ref{lim}, we know that $$\lim _{\rho \rightarrow \infty} {\bf Q}(\varepsilon, \rho)={\bf Q}$$ exists and is independent on $\varepsilon$. Then Lemma \ref{IWI1} shows that ${\bf Q}$ is positive semi-definite. Same as \cite{WXZ}, we can prove that the trace $tr{\bf Q}$, sum of the second-order minors ${\bf Q} ^{(2)}$, sum of the third-order minors ${\bf Q} ^{(3)}$ and the determinant $\det{\bf Q}$ are independent on $t$ as well as on specific Clifford representation. Their nonnegativity gives rise to (\ref{ineq}).

If $E_0=0$ for some end, then by Proposition \ref{Dirac2}, there exists $\{\phi_\alpha\}$ which forms a basis of the spinor bundle over $M$ everywhere such that
\beq
\lim _{\rho \rightarrow \infty} \int_M \chi _\rho | \widehat{\nabla}\phi_\alpha | ^2 d\mu=0.
\eeq
Since $ \widehat{\nabla}\phi_\alpha \in L^2$, Lebesgue's dominated convergence theorem implies
\begin{eqnarray}\label{E1}
 \int_M | \widehat{\nabla}\phi_\alpha | ^2 d\mu=0.
\end{eqnarray}
Let $\psi$ be any $H^1$ spinor with compact support set. Since
\begin{eqnarray}\label{E2}
\begin{aligned}
 &\int_M \langle\widehat{\nabla}_i\widehat{\nabla}_j \phi, \psi \rangle d\mu\\
 =&-\int_M  \langle\widehat{\nabla}_j \phi , \nabla_i\psi \rangle d\mu
-\int_M  \langle\widehat{\nabla}_j \phi , \frac{1}{2}h_{ij}e_0\cdot e_j\cdot\psi \rangle d\mu\\
 &+\int_M \langle\widehat{\nabla}_j \phi , \frac{\sqrt{-1}}{2}\kappa e_i\cdot\psi \rangle d\mu.
\end{aligned}
\end{eqnarray}
By (\ref{E1}) and H\"{o}lder inequality, we know that all the terms in the right hand side of (\ref{E2}) are zero for $\phi=\phi_\alpha$. This implies, in the sense of distribution,
\begin{eqnarray}\label{E3}
 \widehat{\nabla}_i\widehat{\nabla}_j \phi _\alpha=0.
\end{eqnarray}
Therefore, in the sense of distribution,
\begin{eqnarray}\label{E4}
\begin{aligned}
\big(\widehat{\nabla}_i\widehat{\nabla}_j-\widehat{\nabla}_j\widehat{\nabla}_i\big)\phi _\alpha
=\big(\mathcal{P}_1 +\mathcal{P}_2 +\mathcal{P}_3 \big)\phi _\alpha,
\end{aligned}
\end{eqnarray}
where
\begin{eqnarray*}
\begin{aligned}
&\mathcal{P} _{1}=-\frac{1}{4}\big(R_{ijkl}+\kappa^2(\delta_{ik}\delta_{jl}-\delta_{il}\delta_{jk})+h_{ik}h_{jl}-h_{il}h_{jk}\big)e_k\cdot e_l\cdot,\\
&\mathcal{P} _{2}=-\frac{1}{2}(\nabla_i h_{jk}-\nabla_j h_{ik}) e_0\cdot e_k\cdot,\\
&\mathcal{P} _{3}=-\sqrt{-1}\kappa h_{jk}
 e_0\cdot e_i\cdot e_k\cdot+\sqrt{-1}\kappa h_{ik}
 e_0\cdot e_j\cdot e_k\cdot.
\end{aligned}
\end{eqnarray*}
Therefore, by (\ref{E3}), (\ref{E4}), we obtain
\begin{eqnarray*}
\begin{aligned}
2\int_M\chi\langle\P_{1}\phi_\alpha, \phi _\beta\rangle d\mu =&
\int_M\chi\langle(\P_{1}+\P_{2}+\P_{3})\phi_\alpha, \phi _\beta\rangle d\mu \\
- &\int_M\chi\langle\phi _\alpha, (\P_{1}+\P_{2}+\P_{3})\phi _\beta\rangle d\mu =0
\end{aligned}
\end{eqnarray*}
for any for any $\alpha$, $\beta$ and any $H^1$ function $\chi$ with compact support set. In terms of Proposition \ref{Dirac2},
it implies
\beQ
R_{ijkl}=- \kappa^2(\delta_{ik}\delta_{jl}-\delta_{il}\delta_{jk})-h_{ik}h_{jl}+h_{il}h_{jk} \label{gauss}
\eeQ
in the sense of distribution. Furthermore, we get
\beQ
\P_{2}+\P_{3}=0
\eeQ
in the sense of distribution. Now using
\begin{eqnarray*}
\begin{aligned}
e_0\cdot\P_{2}&=-\P_{2}e_0\cdot,\\
e_0\cdot\P_{3}&=\P_{3}e_0\cdot,
\end{aligned}
\end{eqnarray*}
we obtain
\begin{eqnarray*}
\begin{aligned}
2\int_M\chi\langle e_0\cdot\P_{2}\phi_\alpha,\phi _\beta\rangle d\mu
=&\int_M\chi\langle e_0\cdot(\P_{2}+\P_{3})\phi_\alpha, \phi _\beta\rangle d\mu\\
-&\int_M\chi\langle(\P_{2}+\P_{3})e_0\cdot\phi_\alpha, \phi _\beta\rangle d\mu=0.
\end{aligned}
\end{eqnarray*}
Therefore, we obtain
\beQ
\nabla_i h_{jk}-\nabla_j h_{ik}=0 \label{codazzi}
\eeQ
in the sense of distribution. The proof of the theorem is complete. \qed

\begin{rmk}
Due to Theorem \ref{pet}, we define $\sqrt[4]{\det{\bf Q}}$ as the total rest mass of weak asymptotically anti-de Sitter
initial data sets. This total rest mass was defined for usual asymptotically anti-de Sitter initial data sets in \cite{WXZ}.
\end{rmk}


{\footnotesize {\it Acknowledgement. The work of Y. Wang is supported by Chinese NSF grants 11401168, 11671089 and foundation of He'nan
Educational Committee \# 19A110010.
The work of X. Zhang is supported by Chinese NSF grants 11571345, 11731001, the special foundation for Guangxi Ba Gui Scholars and Junwu Scholars, and HLM, NCMIS, CEMS, HCMS of Chinese Academy of Sciences.}

}


\end{document}